\documentstyle[preprint,floats,prl,aps,epsf]{revtex}

\begin{document}

\preprint{\vbox{\hbox {March 1998} \hbox{IFP-757-UNC} }}
\title{Kaon Spontaneous CP Violation Reevaluated}
\author{\bf Paul H. Frampton and Masayasu Harada}
\address{%
University of North Carolina, Chapel Hill, NC  27599-3255}
\maketitle

\begin{abstract}
The CP parameters $\epsilon$ and $\frac{\epsilon^{'}}{\epsilon}$ are 
calculated in the aspon model of spontaneous
CP violation, a model which solves the strong CP problem.
A new range for the scale of spontaneous breaking of CP
is found. It is shown that  $\frac{\epsilon^{'}}{\epsilon}$ 
is suppressed by $\sim x^2 v^2/(\kappa^2 \sin^5 \theta_C) 
\sim  5 \times 10^{-3}$ relative to the Standard Model. If experiment finds that 
$\frac{\epsilon^{'}}{\epsilon}$ is $10^{-4}$ or greater in
magnitude, it will mean that the present approach to
spontaneous CP violation is excluded.
\end{abstract}

\newpage

The origin of CP violation is still not well understood. 
It could arise from explicit breaking, for example 
in the KM mechanism~\cite{KM} of quark flavor mixing of 
three families. Alternatively, it may arise from spontaneous 
CP breaking, for example as in the aspon model~\cite{FK}. This 
model solves the strong CP problem and provides a mechanism 
for weak CP violation which can explain the parameter 
$\epsilon_K$ in the kaon system~\cite{FN}. It also predicts very 
small CP asymmetries in $B^0-\bar{B}^0$ decays~\cite{AFKL} and
production of exotic particles at the LHC\cite{FNKW}.

The purpose of this article is to reevaluate 
the predictions of the aspon model for the 
$K^0$ system, particularly $\epsilon$ and 
Re$(\frac{\epsilon^{'}}{\epsilon})$ more carefully 
than has been done before. This will lead to some 
new predictions and constraints on the parameters of the model.

The aspon model possesses a gauge symmetry 
$(SU(3)_c \times SU(2)_L \times U(1)_Y) \times U(1)_X$.
All the particles of the Standard Model with three families, 
including one doublet Higgs scalar, have
aspon charge $X = 0$. The additional states are a 
non-chiral doublet $Q = (U, D)$ of heavy quarks with $X = 1$ and
two complex singlet Higgs scalars $\chi^{\alpha} (\alpha = 1,2)$
with $X = 1$. With this arrangement, the strong CP problem is
solved (given a certain constraint on the parameters) because
at leading-order the quark mass matrix has a real determinant.
The $\chi^{\alpha}$ develop complex VEVs 
$\langle\chi^{\alpha}\rangle = \rho^{\alpha} e^{i \phi^{\alpha}}$
with $(\phi^{1} - \phi^{2}) \neq 0$, thus breaking CP and 
giving the $U(1)_X$ gauge boson
-the ``aspon''- a mass through the Higgs mechanism.

Let the Yukawa couplings of the light to heavy quarks
be written $h_i^{\alpha} q_L^i Q_R \chi^{\alpha}$ and define 
$x_i = \Sigma_{\alpha} h_i^{\alpha} \langle\chi^{\alpha}\rangle / M$ 
where $M$ is the mass of $Q$ then the requirements of strong CP and
naturalness constrain $|x_i|^2$ to be~\cite{FG}
\begin{equation}
3 \times 10^{-5} < |x_i|^2 < 10^{-3} \label{range}
\end{equation}
for each $i = 1,2,3$ (hereafter
we suppress the subscript
and modulus sign on $|x_i|^2 \rightarrow x^2$).

Let us first evaluate $\epsilon_K$ given by:
\begin{equation}
|\epsilon_K| = \frac{1}{\sqrt{2} \Delta m_K}
\left( Im M_{12} + 2\left(
\frac{Im A_0}{Re A_0} \right) Re M_{12} \right)
\end{equation}
The second term is an order of magnitude smaller 
than the first because  
$(Im A_0)/(Re A_0)$ is orders 
of magnitude less than $|\epsilon_K|$ and $ReM_{12} \sim \Delta m_K$.
The first term then gives, with $\Delta m_K = 3.5 \times 10^{-15}$GeV
and the tree-level aspon exchange (Fig.~1.)
\begin{equation}
|\epsilon_K| = \frac{1}{\sqrt{2} \Delta m_K} Im(x_1^* x_2)^2 
\frac{f_K^2}{3} m_K \frac{2}{\kappa^2}
\end{equation}
with $\kappa = \sqrt{(\rho_1^2+\rho_2^2)}$. Writing 
$Im(x_1^{*}x_2)^2 \rightarrow x^4$ and
using $|\epsilon_K| = 2.26 \times 10^{-3}$ gives the relationship
\begin{equation}
\kappa / x^2 = 2.9 \times 10^7 GeV. \label{constraint}
\end{equation}

Thus the symmetry breaking scale $\kappa$, given 
the range of $x^2$ in Eq.~(\ref{range}) satisfies\cite{FOOTNOTE}
\begin{equation}
29 TeV > \kappa > 870 GeV.
\end{equation}
The aspon mass $M_A = g_A \kappa$ may be estimated, 
taking {\it e.g.} $g_A = 0.3 (=e)$ as $8.7TeV > M_A > 260GeV$.

To evaluate 
$Re(\frac{\epsilon^{'}}{\epsilon})$ 
requires
the study of several Feynman diagrams\cite{SM}, and their comparison to
the Standard Model. Recall that the most recent evaluations in runs at CERN
(NA31)\cite{CERN} and FNAL (E731)\cite{FNAL} give the results 
$Re(\frac{\epsilon^{'}}{\epsilon}) = (23 \pm 3.6 \pm 5.4) \times 10^{-4}$
and 
$Re(\frac{\epsilon^{'}}{\epsilon}) = (7.4 \pm 5.2 \pm 2.9) \times 10^{-4}$
respectively, where the first error is statistical and the second is systematic.
These results are consistent within two standard deviations; the error
is expected to be reduced to $1 \times 10^{-4}$ in foreseeable future
experiments.  
 
We first consider the two tree diagrams shown in
Fig.~2.  An estimate of Fig.~2(a) is $(x^4 v^2 / \kappa^2) \leq 7
\times 10^{-11}$, where we use Eq.~(\ref{constraint}), to be compared
with $\frac{\alpha_s}{4 \pi} (\sin\theta_C)^5  \simeq 10^{-5}$ for
the largest (gluon penguin) Standard Model contribution.
The tree diagram of Fig.~2(b) can be made real by
phase rotations of the quark fields.

Also contributing to $Re(\frac{\epsilon^{'}}{\epsilon})$ at one
loop level are the penguin diagrams of Fig.~3, and the box diagrams of
Fig.~4.

Beginning with the penguins in the Standard Model, the gluon penguin
(Fig.~3(a)) is dominated by charm because $Im(V_{ud}^*V_{us}) = 0$ and
because $Im(V_{cd}^*V_{cs}) \simeq Im(V_{td}^*V_{ts}) \sim
\sin^5\theta_C$ 
while the Feynman amplitude is an order of magnitude
larger for $m_c$ than $m_t$. On the other
hand, the electroweak Z-penguin (Fig.~3(b)) is
proportional to $m_q^2$ and is dominated by top; again it is
$\sim \sin^5\theta_C$ and tends to cancel the gluon penguin\cite{EW}.
Of course, Figs.~3(c) and 3(d) do not exist in the Standard Model.

In the aspon model, the imaginary parts of the penguin
diagrams arise quite differently
from in the Standard Model, because the CKM matrix 
elements are replaced by new expressions at 
the vertices, for example:
\begin{equation}
Im(V_{us}^*V_{ud}) = 0
\end{equation}
\begin{equation}
Im(V_{cs}^*V_{cd}) \sim -A^2\rho (1 - \rho)x^2\sin^6\theta_C    
\end{equation}
\begin{equation}
Im(V_{ts}^*V_{td}) \sim + A^2\rho (1 - \rho)x^2 \sin^6\theta_C  
\end{equation}
\begin{equation}
Im(V_{Us}^*V_{Ud}) = O(x^4)  
\end{equation}
As a consequence, the gluon penguin (Fig.~3(a)) is again dominated by
charm while the Z-penguin (Fig.~3(b)) is dominated by top.
But the replacement of the usual CKM matrix elements 
means a suppression relative to the Standard
Model by a factor $x^2\sin\theta_C \leq 2 \times 10^{-4}$. 
We expect a partial cancellation between the gluon and Z penguins
similar to that in the Standard Model.

For the penguin diagrams peculiar to the aspon model
(Figs.~3(c) and 3(d)), we can dismiss 
the diagram of Fig.~3(d) as negligible, being of order $x^4$. 
On the other hand, the new gluon penguin
of Fig.~3(c) gives a contribution of order
$x^2v^2/\kappa^2$ so that its suppression
relative to the Standard Model gluon penguin
is parametrized by $x^2v^2/(\kappa^2\sin^5\theta_C) 
\leq 5 \times 10^{-3}$.
We find that this diagram (Fig.~3(c)) is
therefore the largest contributor to $Re( \frac{\epsilon^{'}}
{\epsilon})$ in the aspon model. At the same time, we see that 
$Re( \frac{\epsilon^{'}}{\epsilon})$ is highly suppressed, with a magnitude
$\leq 10^{-5}$.

Finally, there are the box diagrams of Fig.~4,
where Fig.~4(c) is peculiar to the aspon
model. This last figure, Fig.~4(c), is actually
proportional to $x^4$ and hence negligible.

In the Standard Model the box diagrams Figs.~4(a) and 4(b)
are smaller than the penguin amplitudes of Figs.~3(a) and 3(b)
in their contribution to $Re( \frac{\epsilon^{'}}{\epsilon})$
because of the interplay between the CKM elements and the masses
$m_q$ interior to the diagram. The Feynman amplitude
$\sim (m_q/m_W)^2$ for the same quark on each
side and $\sim (m_{q_1}/m_W)^2 ln (m_{q_2}/m_{q_1})$
for $m_{q_2} > m_{q_1}$ with different quarks on the two sides.
In the aspon model, it is straightforward to see that for 
similar reasons the box diagrams do not compete with the penguins.

In summary, we have found that the
symmetry-breaking scale $\kappa$ for
spontaneous CP violation
should satisfy $29TeV > \kappa > 870GeV$.
It has also been concluded that
$|Re( \frac{\epsilon^{'}}{\epsilon})| \leq 10^{-5}$
in this model. While
$Re( \frac{\epsilon{'}}{\epsilon})$
is not expected to vanish identically,
it does correspond closely to the superweak model prediction\cite{W}.

Discovery experimentally of
$|Re( \frac{\epsilon^{'}}{\epsilon})| > 10^{-4}$
would certainly mean that this type of approach
is ruled out.

\bigskip
\bigskip
\bigskip
\bigskip

This work was supported in part by the US Department of Energy under Grant
No. DE-FG05-85ER-40219.

\newpage

\begin{figure}
\begin{center}
\ \epsfbox{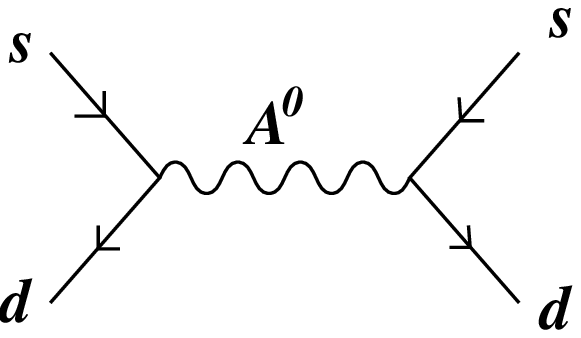}
\end{center}
\caption[]{Tree level aspon ($A^0$) exchange contribution to
$K^0$--$\bar{K}^0$ mixing.}
\end{figure}

\begin{figure}
\begin{center}
\ \epsfbox{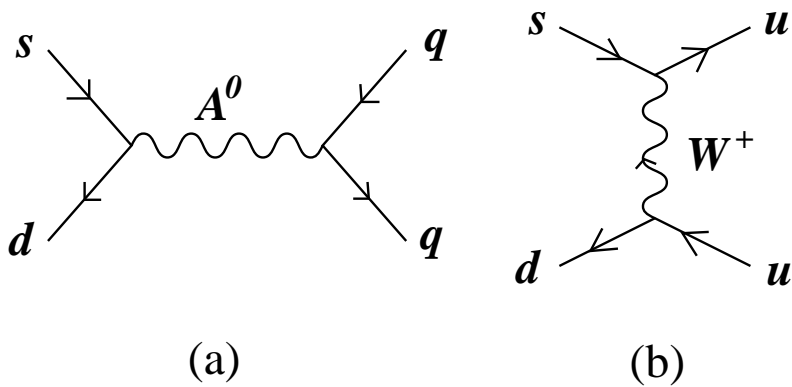}
\end{center}
\caption[]{Tree level contributions to $\epsilon'/\epsilon$:
(a) aspon exchange and (b) $W$ exchange.}
\end{figure}

\begin{figure}
\begin{center}
\ \epsfbox{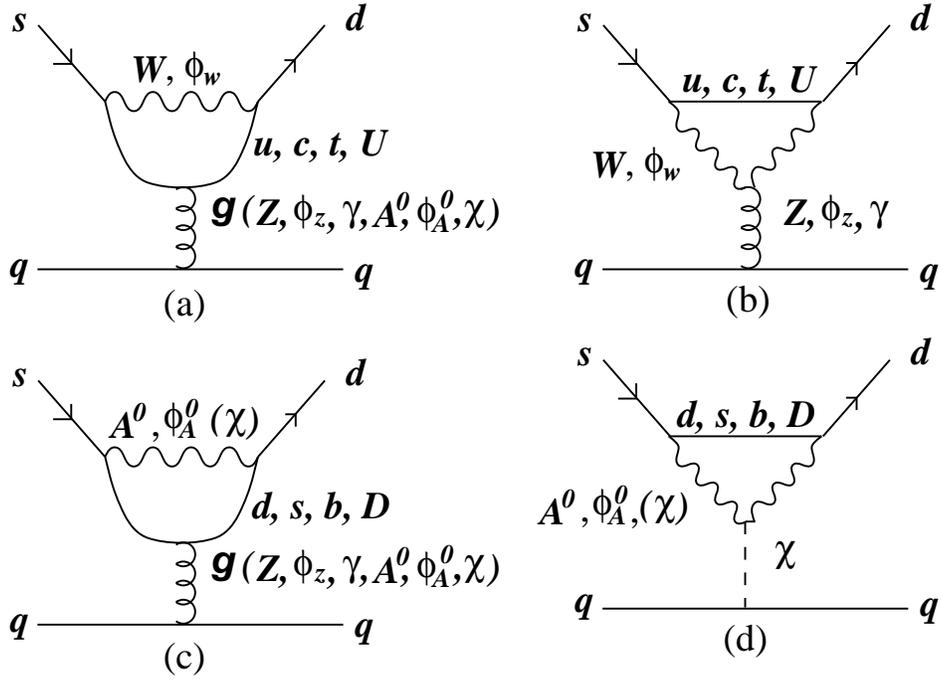}
\end{center}
\caption[]{Penguin diagram contributions to $\epsilon'/\epsilon$.
$\phi_A^0$, $\phi_W$ and $\phi_Z$ are the would-be Nambu-Goldstone
bosons absorbed into the aspon ($A^0$), $W$ and $Z$, respectively.
$\chi$ denotes the massive scalar bosons with aspon charge. 
scalar bosons.}
\end{figure}

\begin{figure}
\begin{center}
\ \epsfbox{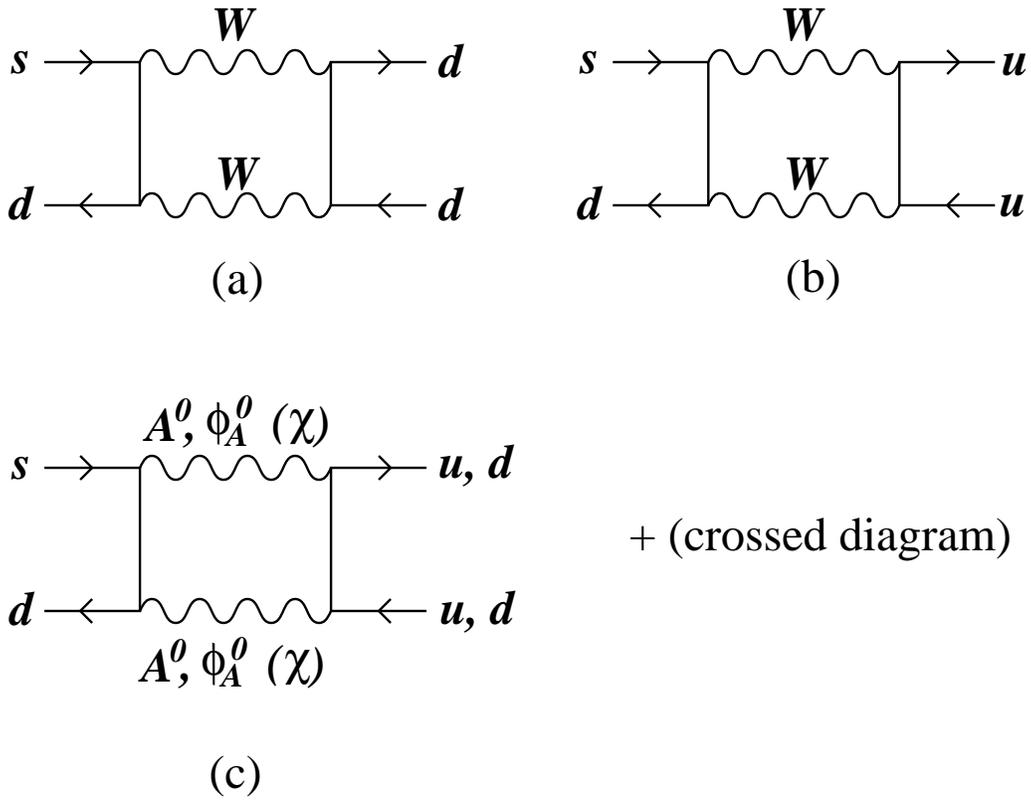}
\end{center}
\caption[]{Box diagram contributions to $\epsilon'/\epsilon$.}
\end{figure}


\begin{thebibliography}{99}
\bibitem{FOOTNOTE}
We have corrected an error in the evaluation of $Im\lambda_{u,t,U}$
following Eq.(33) in \cite{FN}
\end{thebibliography}
\end{document}